\documentclass[preprint,12pt]{elsarticle}
\usepackage{amsmath,amssymb,amsthm,amsfonts,graphics,graphicx,tabularx,array,caption}
\usepackage{booktabs}
\usepackage{multirow}
 
\usepackage{dcolumn}
\usepackage{subcaption}
\usepackage{comment}
\usepackage{bm}
\usepackage{natbib} 
\usepackage[colorlinks=true,
            citecolor=black,
            linkcolor=blue,
            urlcolor=blue]{hyperref}
\newtheorem{definition}{Definition}
\usepackage{etoolbox}

\makeatletter
\patchcmd{\pprintMaketitle}{\Large\@title\par\vskip18pt}{{\Large\bfseries\@title\par}\vskip14pt}{}{}
\patchcmd{\pprintMaketitle}{\elsaddress\par\vskip36pt}{\elsaddress\par\vskip18pt}{}{}
\makeatother
\usepackage[normalem]{ulem}
\newcommand{\rev}[1]{\textcolor{black}{#1}}

\begin{document}
    
\title{AGORA cities: the hidden infrastructure\\ shaping urban fragility}
 
\author[fbk]{Eleonora Andreotti\corref{cor1}}
\ead{eandreotti@fbk.eu}
\author[fbk]{Munazza Usmani}
\author[fbk,unitn]{Alice Barsanti}
\author[fbk]{Maurizio~Napolitano}
 
\cortext[cor1]{Corresponding author.}
 
\affiliation[fbk]{organization={Fondazione Bruno Kessler},
            addressline={Via Sommarive 18},
            postcode={38123},
            city={Povo (TN)},
            country={Italy}}
\affiliation[unitn]{organization={University of Trento},
            city={Povo (TN)},
            country={Italy}}

\begin{abstract}
\rev{Urban services are unevenly distributed across the city, yet it is unclear when their joint presence signals fragility and when it instead outlines a neighbourhood's vocation. For Bologna and Florence, we integrate Foursquare and OpenStreetMap points of interest with Sentinel-2-derived green indicators based on the 3–30–300 rule, cluster urban areas accordingly, and benchmark the clusters against the demographic, social, economic and composite fragility indices produced by the municipal statistical offices, which are exogenous to the analysis. In both cities, we identify a subset of infrastructures, termed AGORA (Arts, Green environments, and Open Routes for Access and mobility), whose joint provision is associated with lower fragility. The AGORA Alignment score (AA), based on the rank correlation between AGORA endowment and composite fragility, indicates full alignment in Florence (AA = 1.00) and strong but incomplete alignment in Bologna (AA = 0.80). The remaining service categories instead show wide and heterogeneous distributions: they differentiate areas without tracking fragility and can be interpreted as expressions of urban vocations. The least fragile cluster, the only one with full AGORA endowment, hosts just 3.3\% of residents in Bologna and 6.3\% in Florence. The framework offers a procedure transferable to other contexts and identifies the expansion of AGORA infrastructures as a candidate lever for reducing fragility without altering existing vocations.}
\end{abstract}


\maketitle

\section{Introduction}
Cities are inherently complex systems \citep{batty2009cities, barthelemy2016structure}, shaped by dynamic interactions among social, economic, environmental, and infrastructural factors \citep{Batty2008, Bettencourt2013, alessandretti2020scales, andreotti2025scalefree}, and where spatial and functional patterns emerge through self-organizing dynamics driven by cumulative and decentralized processes \citep{Portugali2000SelfOrganization, Batty2007CitiesComplexity}. Such emergent structures resemble what Alexander defines as natural cities, characterized by overlapping and adaptive functions, in contrast to artificial cities shaped by rigid hierarchical planning \citep{Alexander1965CityNotTree}, and they are manifested in the heterogeneous nature of urban areas, where neighborhoods exhibit distinct characteristics, functions, and challenges \citep{LIU2025103807}.
\rev{Some areas benefit from services, green spaces, and accessibility, while others face infrastructural deficiencies, limited public amenities, and environmental degradation \citep{kaufmann2022scaling, NambazoNazombe2024, Bakhtsiyarava2024}. In European cities, socially disadvantaged groups are less served by the cooling benefits of urban green spaces \citep{Rocha2024}, and the value of greenery depends on the infrastructural quality and recreational functions that accompany it \citep{Bottero2022UrbanParks}.}\\
\rev{The relationship between greening and socioeconomic conditions runs in both directions. Urban expansion drives vegetation loss while indirect socioeconomic factors can promote greening, with divergent outcomes between cities of the Global North and Global South \citep{Chen2025Contrasting}; conversely, earlier greening helps explain subsequent gentrification in 17 of 28 cities across Western Europe and North America, a variation best accounted for at the city level and moderated by housing affordability and anti-displacement policies \citep{Anguelovski2022green}. In both directions, outcomes depend on local policy and institutional context, even among structurally comparable cities, which calls for analysing green assets jointly with service distribution and local fragility.}\\
\rev{Throughout this study we use fragility to denote the multidimensional disadvantage experienced by populations living in precarious demographic, social and economic conditions, a construct developed within the distributive strand of the environmental-justice literature and operationalised by municipal authorities to identify areas requiring targeted intervention. We measure it through the construct by which Italian municipal statistical offices quantify this condition at small-area level: a composite of eighteen demographic, social and economic indicators aggregated through a constrained Adjusted Mazziotta--Pareto Index \citep{MazziottaPareto2016}, as implemented for Bologna and Florence \citep{ComuneBolognaFragilita2024, RegioneToscanaFragilitaFirenze2023}, detailed in Table~S1 in the Supplementary Material (SM).}\\
\rev{Our aim is to identify which categories of urban infrastructures and services are associated with fragility. To this end, we map services ranging from commercial activities to healthcare and transportation through Foursquare \citep{foursquare} and OpenStreetMap \citep{osm} points of interest (POIs), harmonised following Soulas et al.~\citep{Lilou2025}, and assess green infrastructure through the 3--30--300 rule \citep{Konijnendijk2023}: every resident should be able to see at least 3 trees from their home, live in a neighbourhood with at least 30\% tree canopy cover, and be no more than 300~m from the nearest green space. All service categories and green indicators enter a clustering of urban areas, whose results are benchmarked against the municipal fragility indices. Methodologically, the study contributes to urban data science by combining multi-source spatial data and clustering techniques to uncover infrastructural inequities and their links to fragility.}\\
\rev{A specific subset of infrastructures emerges from the data: cultural-community amenities, green and recreational environments, and open mobility services, whose joint presence is associated with lower fragility. We term this configuration AGORA, the emergent civic space where accessibility, exchange, and shared use converge, and refer to urban contexts in which fragility decreases as these infrastructures co-occur as AGORA-Aligned. The remaining categories differentiate areas functionally without tracking fragility; we refer to them as vocational, as they reflect the functional tendencies of each area.}\\
\rev{We focus on Bologna and Florence for two reasons. First, both municipalities have published detailed fragility indicators at the level of local areas \citep{ComuneBolognaFragilita2024,RegioneToscanaFragilitaFirenze2023}, enabling a direct benchmark. Second, the two cities are structurally similar, with comparable population sizes, analogous morphologies shaped by surrounding hills, and similar positions within the central Italian urban system. Applying the same framework to both allows us to assess whether the relationship between infrastructure configurations and fragility is reproduced across comparable cities or reflects context-specific differences.}\\
The remainder of the paper is structured as follows. Section \ref{sec:meth} presents the methodological framework, including the rationale for adopting a finer spatial division in Bologna and Florence, the integration of Foursquare and OpenStreetMap POIs into a unified classification system, the identification of green areas and computation of the 3--30--300 indicator, and the clustering procedure. Section \ref{sec:results} reports the results of the clustering analysis for Bologna and Florence, with a benchmarking exercise against the fragility indices. Finally, Section \ref{sec:discussion} discusses the broader implications of the findings for urban policy and planning, and concludes by summarizing the main results and outlining directions for future research.

\section{Methodology}\label{sec:meth}
\subsection{Rationale for a Finer Spatial Division of Bologna}
\rev{Bologna is currently divided into six large districts — \texttt{Borgo Panigale-Reno}, \texttt{Navile}, \texttt{Porto-Saragozza}, \texttt{San Donato-San Vitale}, \texttt{Santo Stefano}, and \texttt{Savena} (Figure S1 in the SM). This structure, established in 2016, is the result of successive reorganizations that reduced the districts from an original set of eighteen. The consolidation simplifies governance, but it does not reflect the way Bolognese residents refer to their urban environment: the names of the former districts (Bolognina, Corticella, Barca, Lame, San Donato, and others) remain in everyday use, in local media and in political discourse, and correspond to long-standing neighborhoods with distinct social, economic, and architectural identities. We therefore adopt as units of analysis the 90 statistical areas (Figure S2 in the SM), the official subdivisions used by the municipality for demographic and urban analysis, whose granularity is closer to this lived geography than the district level.}

\subsection{Rationale for a Finer Spatial Division of Florence}
Florence is administratively divided into five large districts — \texttt{Centro Storico}, \texttt{Campo di Marte}, \texttt{Gavinana–Galluzzo}, \texttt{Isolotto–Legnaia}, and \texttt{Rifredi} (Figure S3 in the SM). \rev{As} in the case of Bologna, these coarse boundaries do not capture the internal heterogeneity of the urban fabric or the lived spatial experience of residents. Unlike Bologna, \rev{however,} Florence \rev{has no} historical system of finer administrative divisions that can be reused\rev{: the} former \textit{quartieri} are only four and thus too coarse to serve as analytical units. \rev{We therefore adopt} the 74 \textit{elementary areas} (\textit{aree elementari}) provided by the Municipality of Florence \citep{FirenzeAE2021}, the smallest official statistical units used for socio-demographic reporting (Figure S4 in the SM). Their size \rev{makes} them directly comparable to the 90 statistical areas employed \rev{for} Bologna\rev{, ensuring methodological consistency and supporting} the study of localized phenomena while retaining robustness in terms of data availability.

\subsection{Integration of Foursquare and OpenStreetMap POI Data}
\begin{figure}[htp]
    \centering
    \begin{subfigure}[b]{0.495\linewidth}
        \centering
        \includegraphics[width=\linewidth]{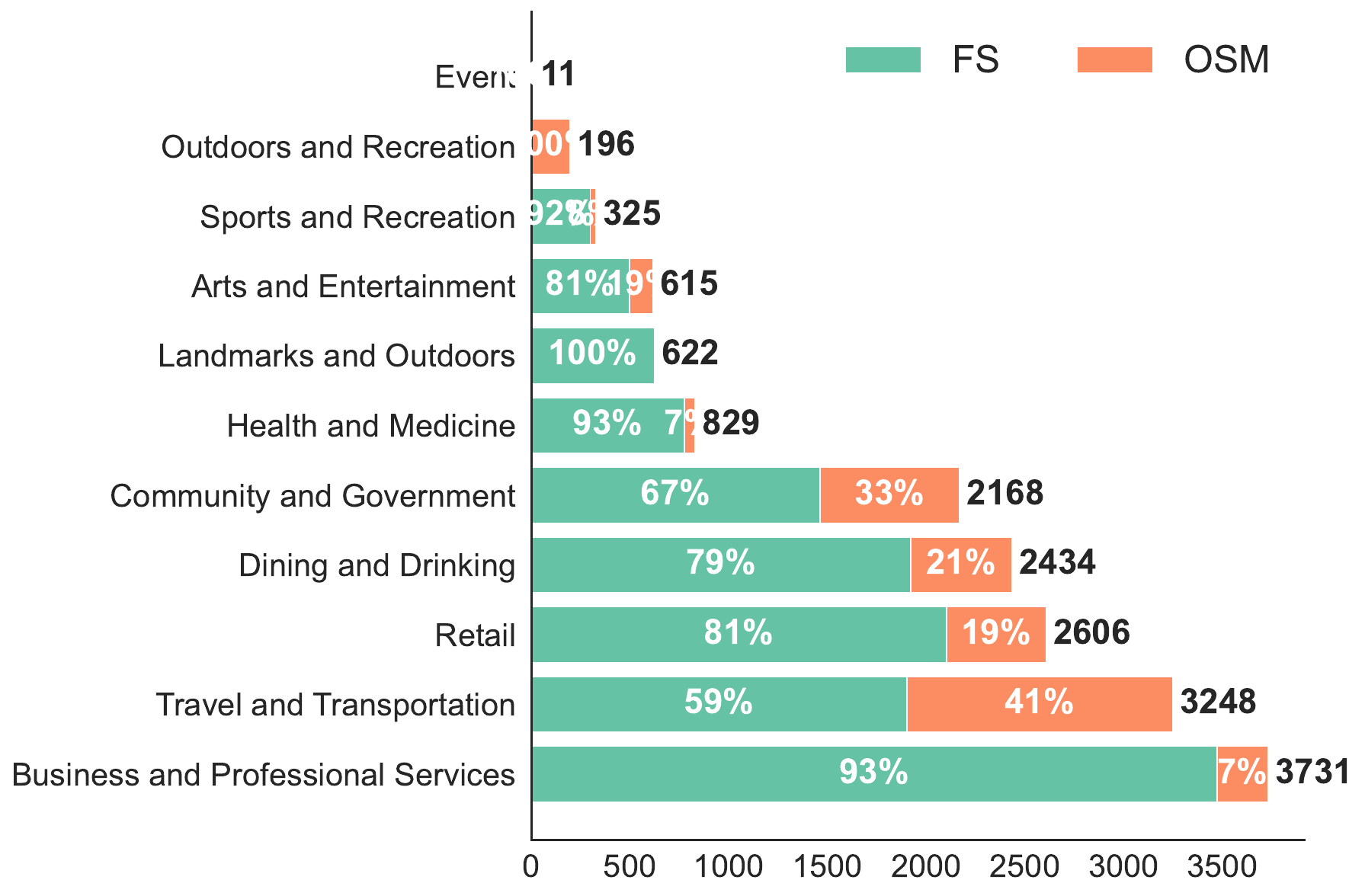}
        \caption{Bologna}
        \label{fig:FSvsOSM_BO}
    \end{subfigure}
    \begin{subfigure}[b]{0.495\linewidth}
        \centering
        \includegraphics[width=\linewidth]{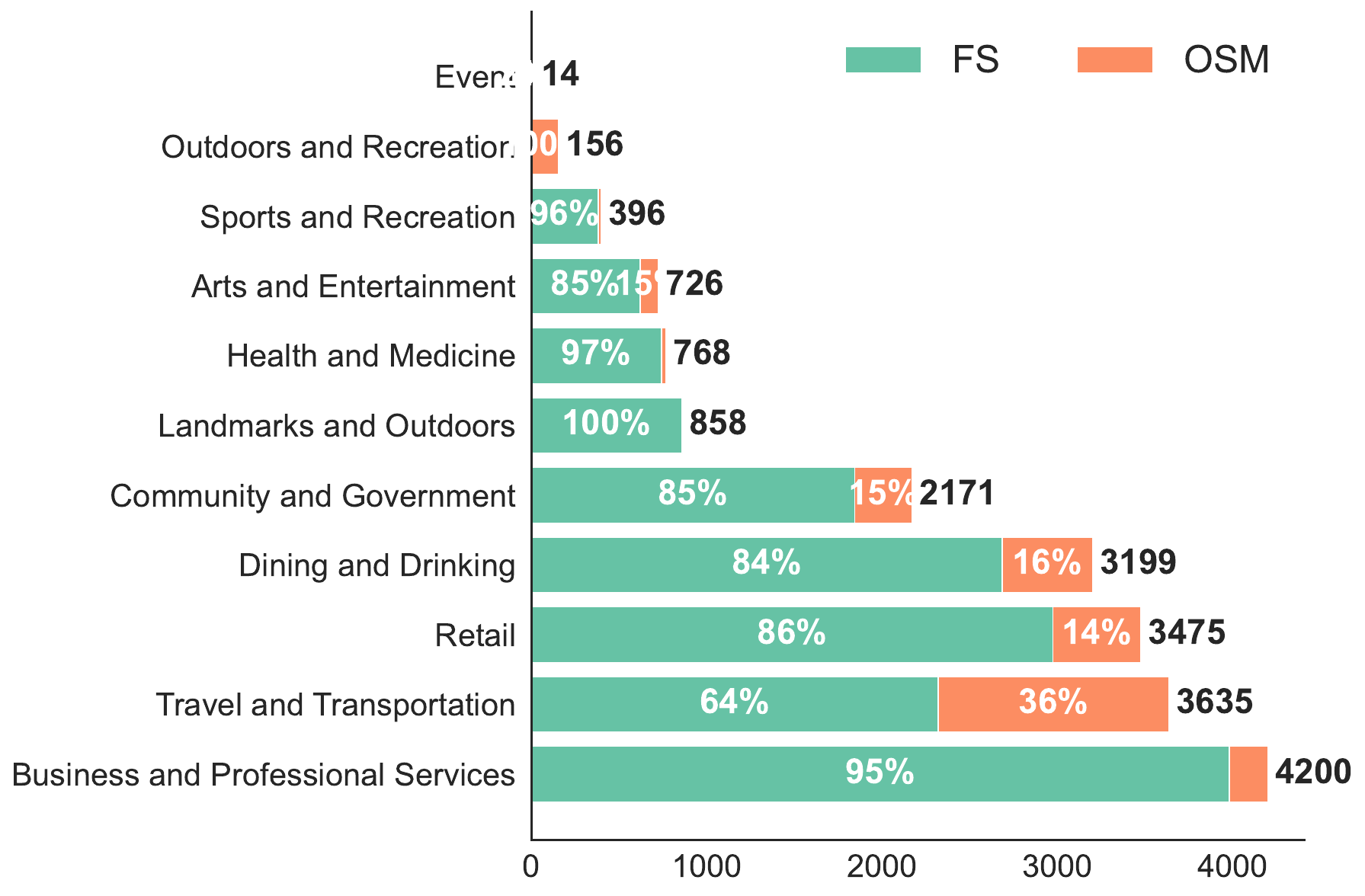}
        \caption{Florence}
        \label{fig:FSvsOSM_FI}
    \end{subfigure}
    \caption{Stacked horizontal bar charts showing the number and percentage of POIs by source (Foursquare and OSM) across Level-1 categories for (a) Bologna and (b) Florence.}
    \label{fig:FSvsOSM}
\end{figure}

\rev{For the vocational analysis, we rely on POIs from Foursquare as the primary data source~\citep{foursquare}, integrating additional POIs from OpenStreetMap (OSM) \citep{osm} reclassified into the same Foursquare taxonomy at the first-level categories, following the mapping method proposed by~\citep{Lilou2025}. Figures~\ref{fig:FSvsOSM_BO} and~\ref{fig:FSvsOSM_FI} present the resulting distribution across Level-1 categories, disaggregated by source, for Bologna and Florence.
POIs serve as a proxy for the vocational profile of urban areas, as they represent tangible facilities, services, and activities that shape everyday urban life. While services such as healthcare or transportation are expected to cover the entire urban territory, categories like \texttt{Dining and Drinking} or \texttt{Retail} are typically concentrated in specific zones, and it is unlikely that the same areas simultaneously exhibit high densities across all categories \citep{andreotti2025scalefree}. Analysing their density and composition therefore makes it possible to identify predominant functions and detect mixed-use areas; the spatial patterns are further explored through kernel density estimation applied separately to each Level-1 category, overlaid on the respective street networks (Figures S5 and S6 in the SM). Both sources are subject to uneven spatial coverage: volunteered geographic information varies in completeness and positional accuracy across space \citep{Haklay2010}, while commercial datasets depend on user activity and may under-represent low-footfall areas. The integration of the two sources is itself our response to this limitation, since their production mechanisms differ and combining them recovers information that either would miss on its own. Residual coverage differences cannot be excluded, and we acknowledge this as a constraint inherent to open data sources.}\\

\subsection{Identification of Green Areas and Computation of the 3--30--300 Indicator}
\begin{figure}[htp]
    \centering
    
    \begin{subfigure}[b]{0.32\linewidth}
        \centering
        \includegraphics[width=\linewidth]{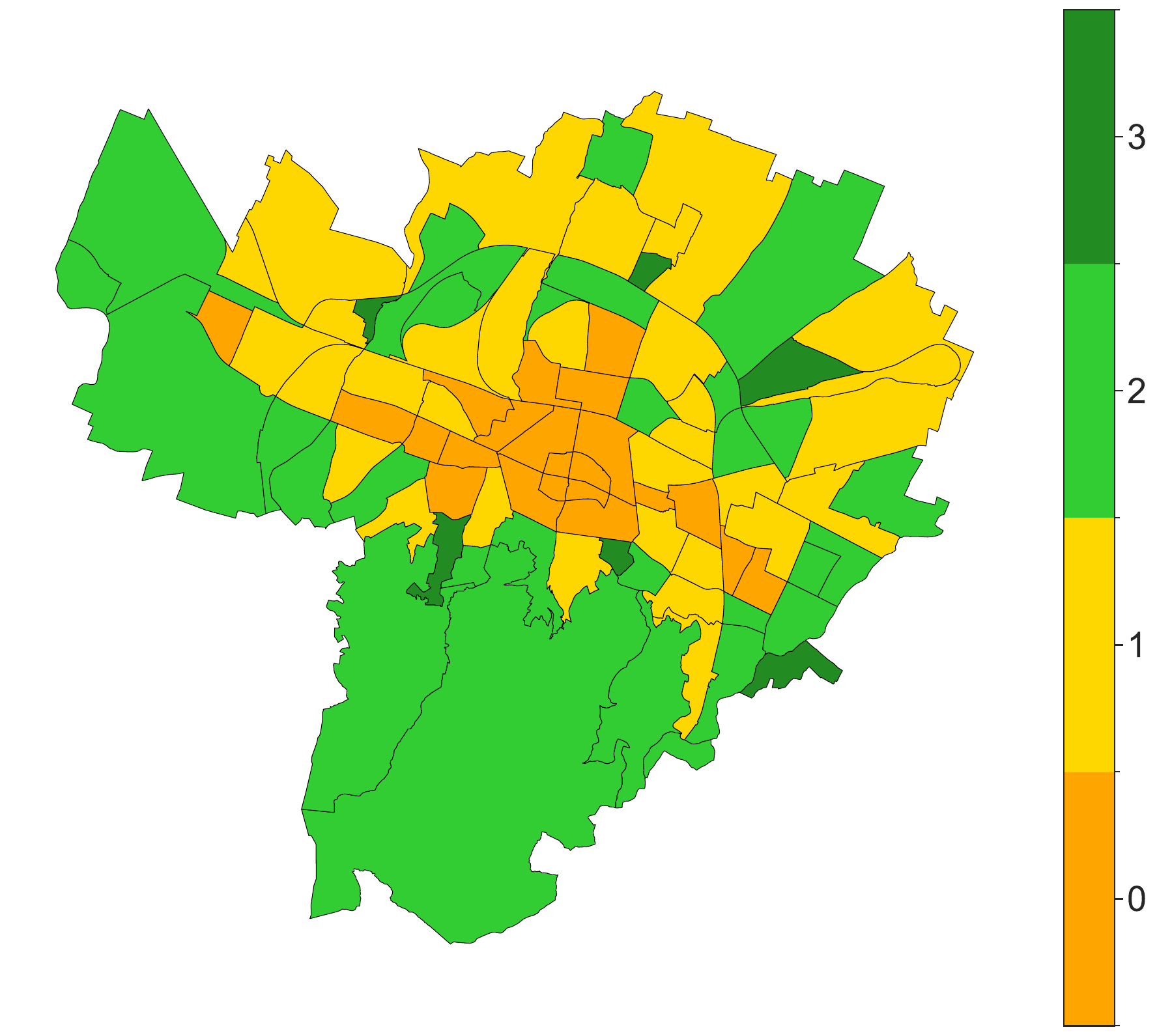}
        \caption{Bologna}
        \label{fig:3-30-300BO}
    \end{subfigure}
     \begin{subfigure}[b]{0.32\linewidth}
        \centering
        \includegraphics[width=\linewidth]{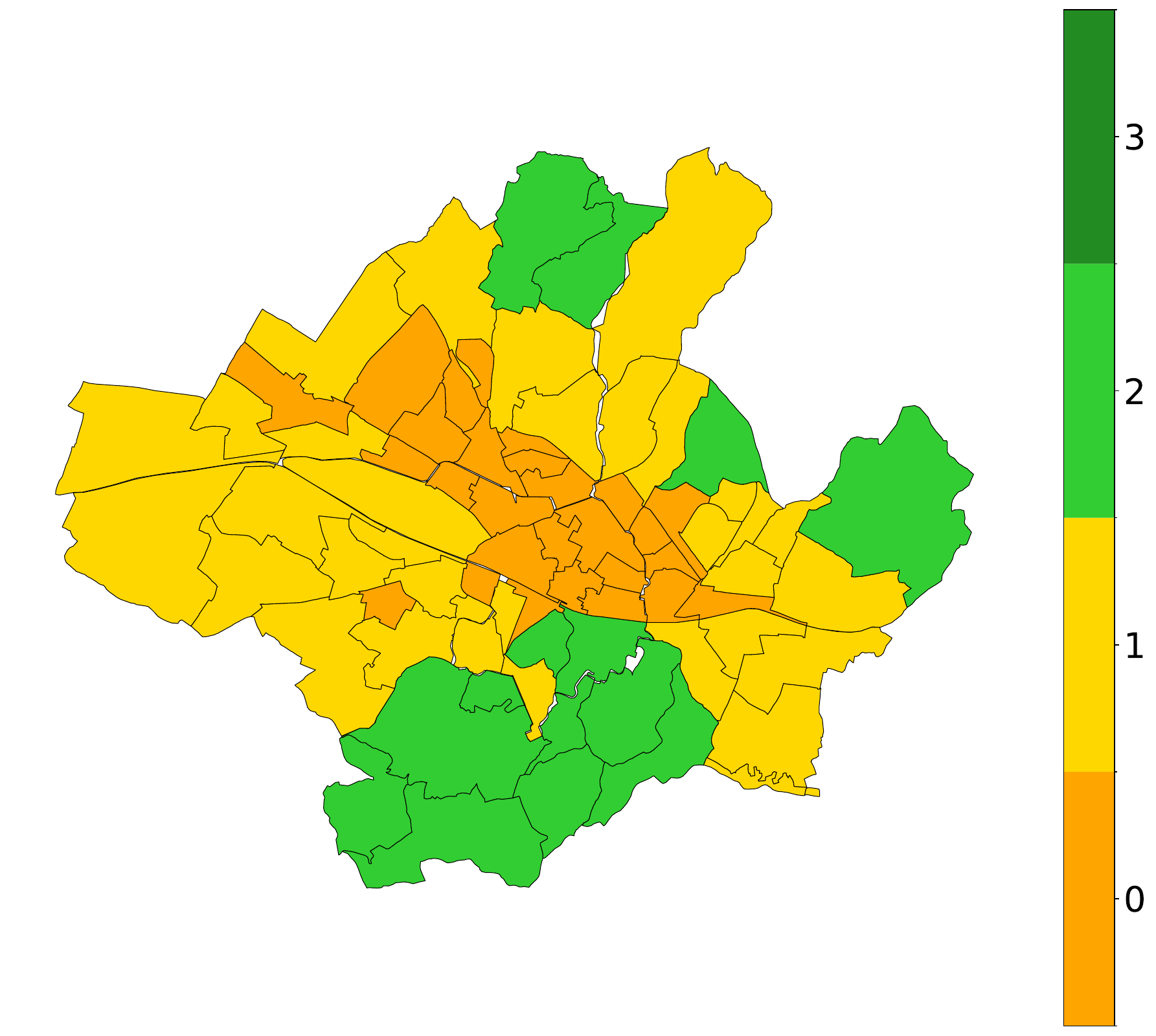}
        \caption{Florence}
        \label{fig:3-30-300FI}
    \end{subfigure}
    \begin{subfigure}[b]{0.32\linewidth}
        \centering
        \includegraphics[width=\linewidth]{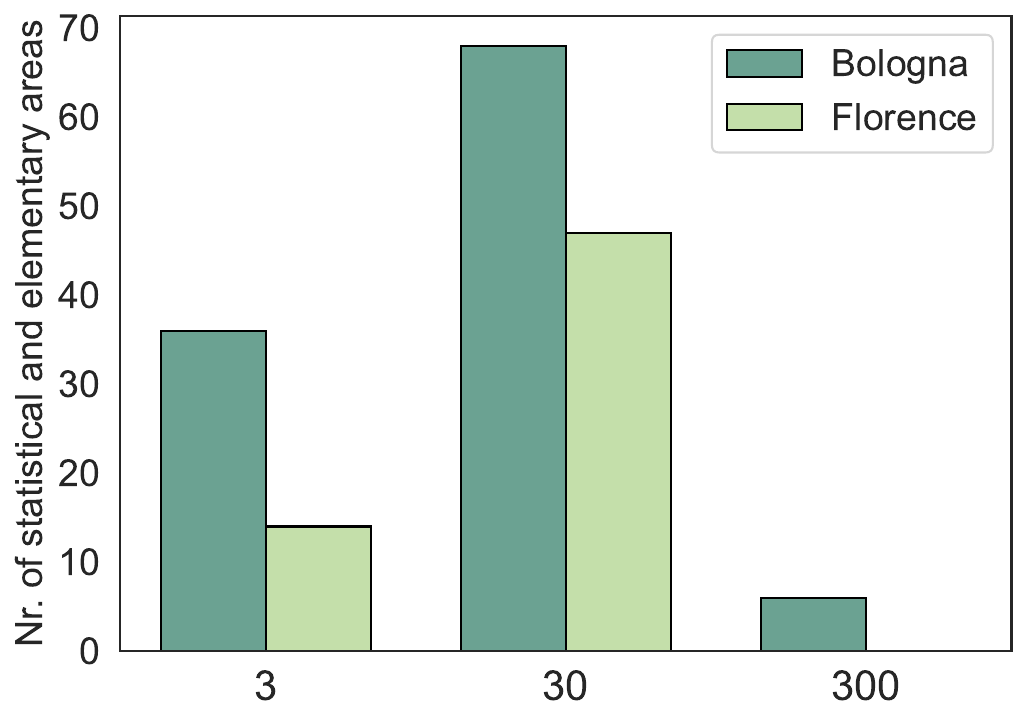}
        \caption{Number of territorial units meeting each of the three 3–30–300 indicators.}
        \label{fig:3-30-300isto}
    \end{subfigure}
    
    \caption{Compliance with the 3–30–300 urban greening framework in Bologna and Florence. Panels (a) and (b) show the number of 3–30–300 conditions met (0–3) across territorial units in Bologna and Florence, respectively. Panel (c) reports the number of units satisfying each individual indicator: proximity to at least three trees per residence (3), tree canopy cover of at least 30\% (30), and proximity within 300 m to a park (300).}
    
    \label{fig:3-30-300}
\end{figure}

The 3--30--300 rule, recently formulated by \citep{Konijnendijk2023}, provides an evidence-based framework for assessing the accessibility and quality of urban green spaces at the residential scale, linking individual well-being with urban forest structure through three complementary targets: every resident should be able to see at least three trees from their home, live in a neighbourhood with at least 30\% tree canopy cover, and have access to a public green space within 300~m. The composite indicator thus captures visibility, coverage, and proximity.\\
The computation relied on two geospatial data sources. The first is a raster layer derived from Sentinel-2 imagery (year 2025) using Google Earth Engine (GEE)\rev{, available at a spatial resolution of 10~m for the bands used here}, filtered by cloud cover (\textless10\%) and clipped to the study area; NDVI was calculated from the NIR (Band 8) and Red (Band 4) bands, and a threshold of NDVI \rev{\textgreater} 0.3 was applied to identify dense vegetation corresponding to tree canopy. This dataset was used both to compute the canopy cover component and to estimate \rev{canopy reference locations}, by converting all pixels with canopy value equal to one into centroid points. The second source consists of polygon features representing public parks and gardens extracted from the official \emph{Database di Base della Statistica Nazionale} (DBSN) \citep{dbsn}, validated to ensure topological correctness. Building footprints were retrieved from OpenStreetMap by filtering residential structures (\texttt{building=residential, apartments, house, detached, yes}) and served as the reference units.\\
The ``3 trees'' component was evaluated by computing the centroid of each building footprint, generating a circular buffer of radius 50~m, and counting the \rev{canopy} points derived from the \rev{Sentinel-2} canopy dataset within it; buildings with at least three \rev{such points} were considered compliant. The ``30\% canopy'' component was assessed at the area level as the proportion of canopy pixels relative to all valid pixels in the same raster. The ``300~m to a park'' component was measured by buffering the DBSN polygons by 300~m and selecting buildings whose footprints intersected the buffered area. \rev{For the building-level components, an area was considered compliant when all its residential buildings met the condition.} A composite score, defined as the number of criteria satisfied, was then computed for each statistical area in Bologna and each elementary area in Florence, with values ranging from zero to three.\\
\rev{Two limitations should be noted. At a resolution of 10~m each pixel corresponds to 100~m$^2$, so canopy pixels cannot be equated with individual trees: our implementation of the ``3 trees'' component adopts the buffer-based distance threshold established in the literature \citep{Bertassello2026}, in which the 50~m radius approximates the range within which trees are plausibly visible from the dwelling, and should therefore be read as a measure of canopy presence in the immediate residential surroundings. In addition, NDVI thresholding captures photosynthetically active vegetation in general and does not discriminate tree canopy from lawns or cultivated land, which may lead to overestimation in hillside areas. Deriving canopy estimates from satellite imagery rather than from public green inventories is nonetheless deliberate, as it captures the private and semi-private greenery that institutional databases omit and that accounts for a substantial share of tree cover in both cities.}\\
Figure~\ref{fig:3-30-300isto} shows the number of territorial units meeting each individual indicator, revealing similar patterns in both cities: the 30\% canopy cover is the most frequently achieved condition, whereas proximity within 300~m to a park is the most challenging. The spatial distribution of the composite results is illustrated in Figure~\ref{fig:3-30-300BO} for Bologna and Figure~\ref{fig:3-30-300FI} for Florence.

\subsection{Clustering Analysis of Urban Areas}
To explore functional similarities among territorial units, we applied an unsupervised K-means clustering \rev{\citep{MacQueen1967, Lloyd1982}} to a feature matrix whose rows are statistical areas in Bologna \rev{and} elementary areas in Florence\rev{, hereafter simply} \textit{areas}. Columns include (i) the number of POIs for each Level-1 POI category and (ii) three green-infrastructure indicators encoded as binary variables (1 if the \rev{area} met the target, 0 otherwise). \rev{POI counts were rescaled to the [0,1] interval through a row-wise min--max normalization, that is, using the minimum and maximum observed across categories within each area. Each area is therefore described by the relative composition of its service profile, by which categories predominate locally, rather than by its absolute endowment, so that no single feature dominates the distance computation and the partition reflects functional profiles rather than size effects. The three green indicators are already defined on the same interval and enter unchanged.}\\
\rev{The algorithm partitions the areas into $k$ groups by minimising the within-cluster sum of squared distances to the cluster centroids, following the iterative assignment and update procedure of \citep{Lloyd1982}. Centroids were initialised using the k-means++ scheme \citep{ArthurVassilvitskii2007}, and the algorithm was run with multiple random restarts, retaining the solution with the lowest within-cluster sum of squares, as implemented in scikit-learn \citep{Pedregosa2011}.}\\
To determine the appropriate number of clusters, we \rev{combined} the elbow method\rev{, which identifies} the point at which additional clusters yield diminishing returns in reduced within-cluster variance, \rev{with} the silhouette score \citep{Rousseeuw1987}, \rev{which evaluates} consistency within clusters and separation between \rev{them. To verify that the resulting partition was not an artefact of the algorithm's stochastic components, the clustering was re-estimated across multiple random seeds with a high number of initialisations, measuring the agreement between the resulting partitions through the Adjusted Rand Index \citep{HubertArabie1985}, and cluster cohesion was evaluated via bootstrap consensus clustering \citep{Monti2003}, in which areas were resampled with replacement and re-clustered, recording the co-assignment frequency of each pair as a measure of partition stability.}

\subsection{Benchmark with Municipal Fragility Indices}

To evaluate the predictive potential of the clustering framework, the results were benchmarked against four fragility indices produced by the municipal statistics offices, encompassing demographic, social, economic, and composite fragility. These indices are computed from 18 analytical variables normalized and aggregated through a constrained Adjusted Mazziotta-Pareto Index (AMPI), covering demographic change, economic disadvantage, and social \rev{fragility} as detailed in Table~S1 in the SM. For Bologna, the indices are provided in \citep{ComuneBolognaFragilita2024} and publicly released at \citep{opendatafrag}, whereas for Florence they are sourced from \citep{RegioneToscanaFragilitaFirenze2023} and disseminated via \citep{OpendataFI2023}.
\rev{The choice of these indices rests on their independence from our analysis: the
eighteen variables and their aggregation are defined by the municipal statistical
offices, and we adopt the resulting measures unchanged. The benchmark is therefore
exogenous to the clustering and cannot be adjusted to it. Among the available
sources, these indices were selected because they are computed from the same
variables through the same procedure in both cities, are released at a spatial
resolution compatible with our units of analysis, and are publicly accessible. 
A further advantage is that these sources provide not only a composite score but
also its constituent demographic, social and economic dimensions. This disaggregation
allows the benchmark to be conducted along distinct axes of fragility, which is
essential because different dimensions may respond to infrastructural endowments in
divergent ways: an area may be economically robust yet demographically fragile, or
vice versa. Examining each dimension separately therefore makes it possible to
identify which population segments are most exposed in a given urban configuration,
and to avoid the loss of information that would follow from relying on the composite
measure alone.}\\
Each index, originally computed for small statistical areas, was subsequently rescaled to the interval [0,1] and spatially matched with the clustering results. \rev{Statistical areas for which the municipality does not compute fragility indices, non-residential functional units such as parks, hospital and research sites, were excluded from the benchmark rather than imputed, since fragility is not defined in the absence of a resident population. These areas are nonetheless retained in the clustering, which relies on service and green indicators that are available for them, so that the partition remains independent of the benchmark.} This comparison allows us to assess whether unsupervised clusters, derived exclusively from service and green-infrastructure indicators, align with official fragility patterns. The benchmark analysis thus serves both as a validation step for the proposed framework and as a diagnostic tool to determine which urban vocations are more closely associated with specific fragility dimensions.


\section{Results}\label{sec:results}
\subsection{Clustering of Statistical Areas in the city of Bologna}
\begin{figure}[!ht]
    \centering
    \begin{subfigure}[b]{0.6\linewidth}
        \centering
        \includegraphics[width=\linewidth]{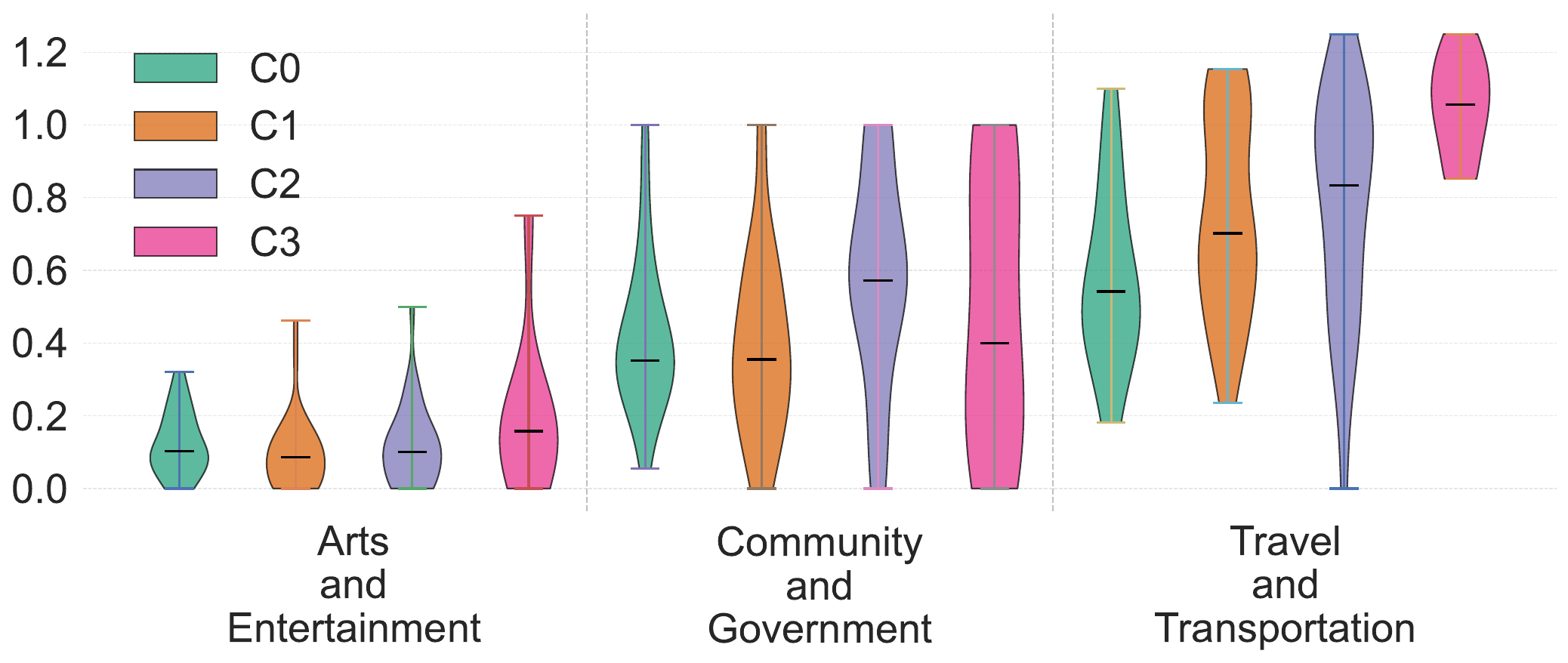}
        \caption{Cultural, governmental, and transportation-related categories.}
        \label{fig:violins_group1}
    \end{subfigure}
    \begin{subfigure}[b]{0.6\linewidth}
        \centering
        \includegraphics[width=\linewidth]{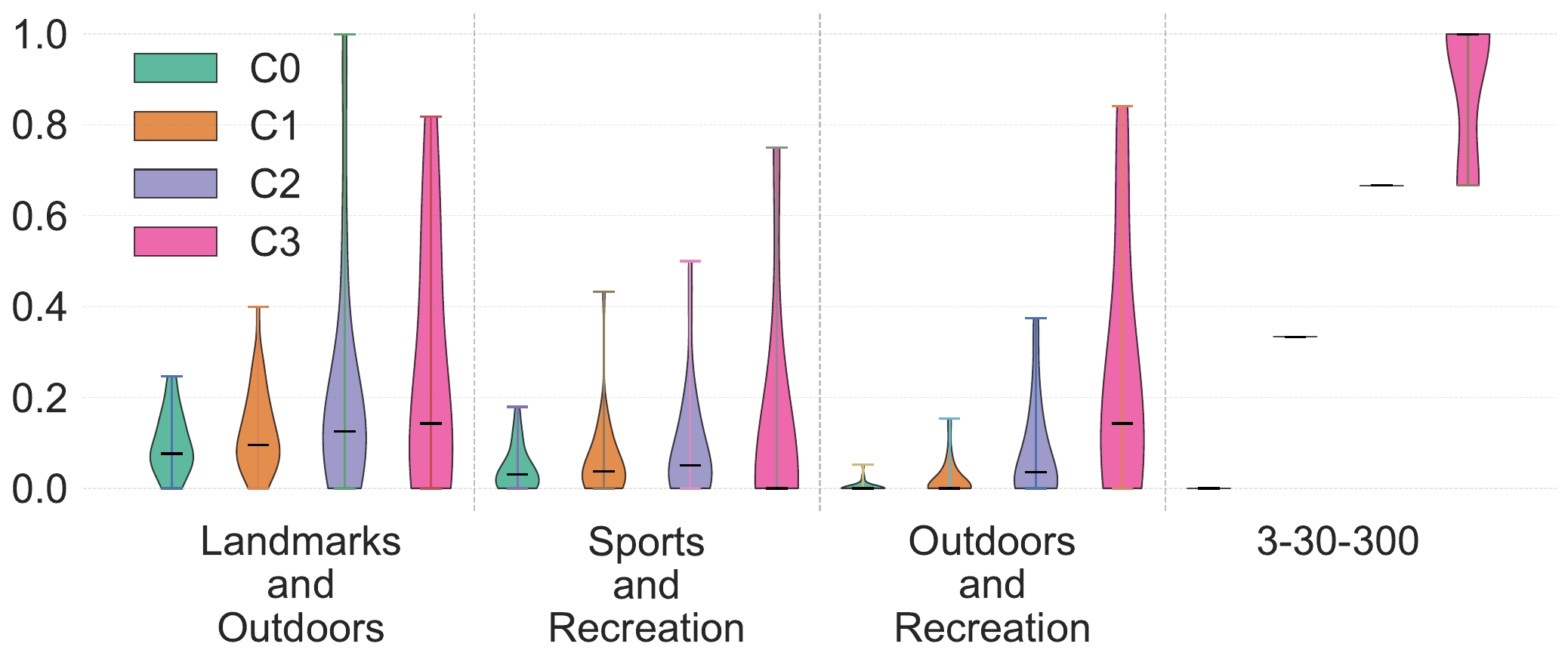}
        \caption{Outdoor, recreational, and green-related functions.}
        \label{fig:violins_group2}
    \end{subfigure}
    \begin{subfigure}[b]{0.6\linewidth}
        \centering
        \includegraphics[width=\linewidth]{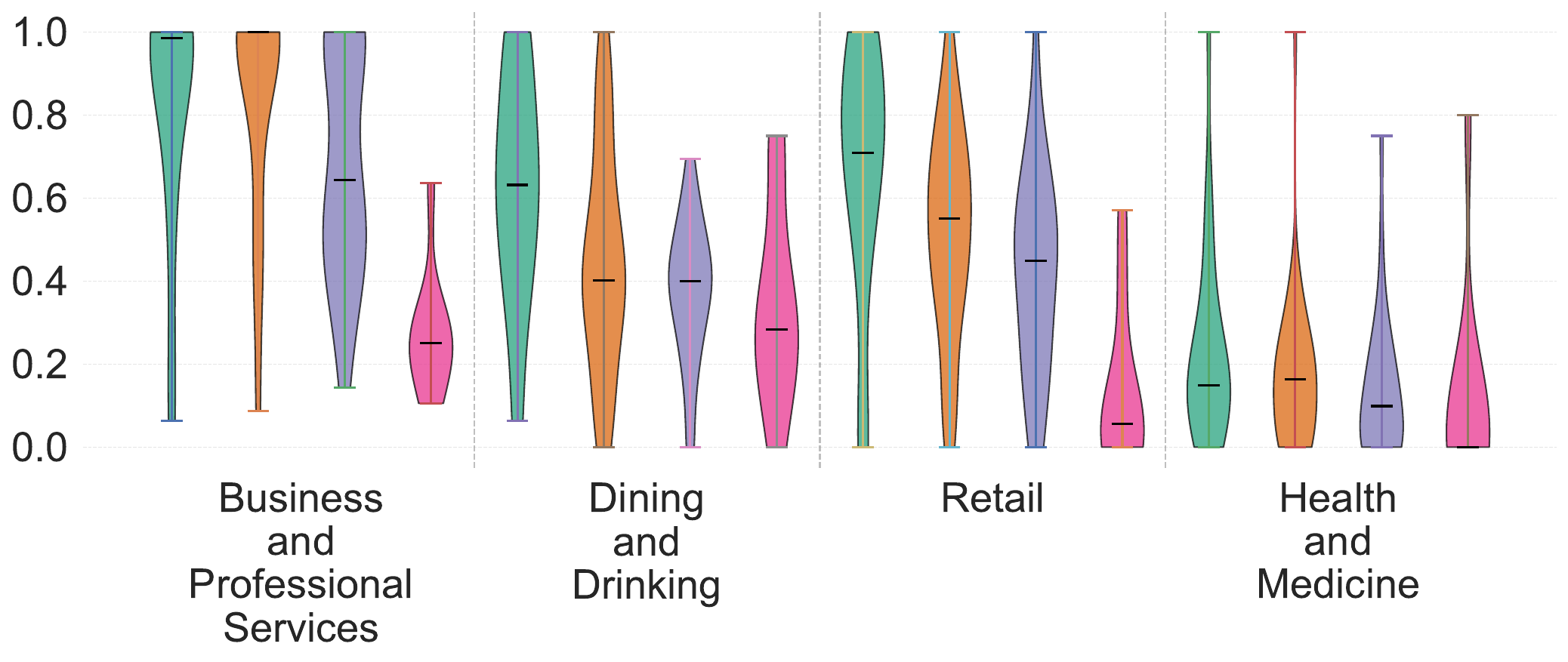}
        \caption{Health, dining, and basic retail services.}
        \label{fig:violins_group3}
    \end{subfigure}
    
    \caption{Cluster structure and functional composition of Bologna's statistical areas. \rev{Panels (a)–(c) report} the distribution of selected Level-1 POI categories across clusters, highlighting differences in cultural, green, and service-oriented functions.}
    
    \label{fig:BO_clusters_full}
\end{figure}

\begin{figure}[!ht]
\centering
\begin{subfigure}[b]{0.32\linewidth}
    \centering
    \includegraphics[width=\linewidth]{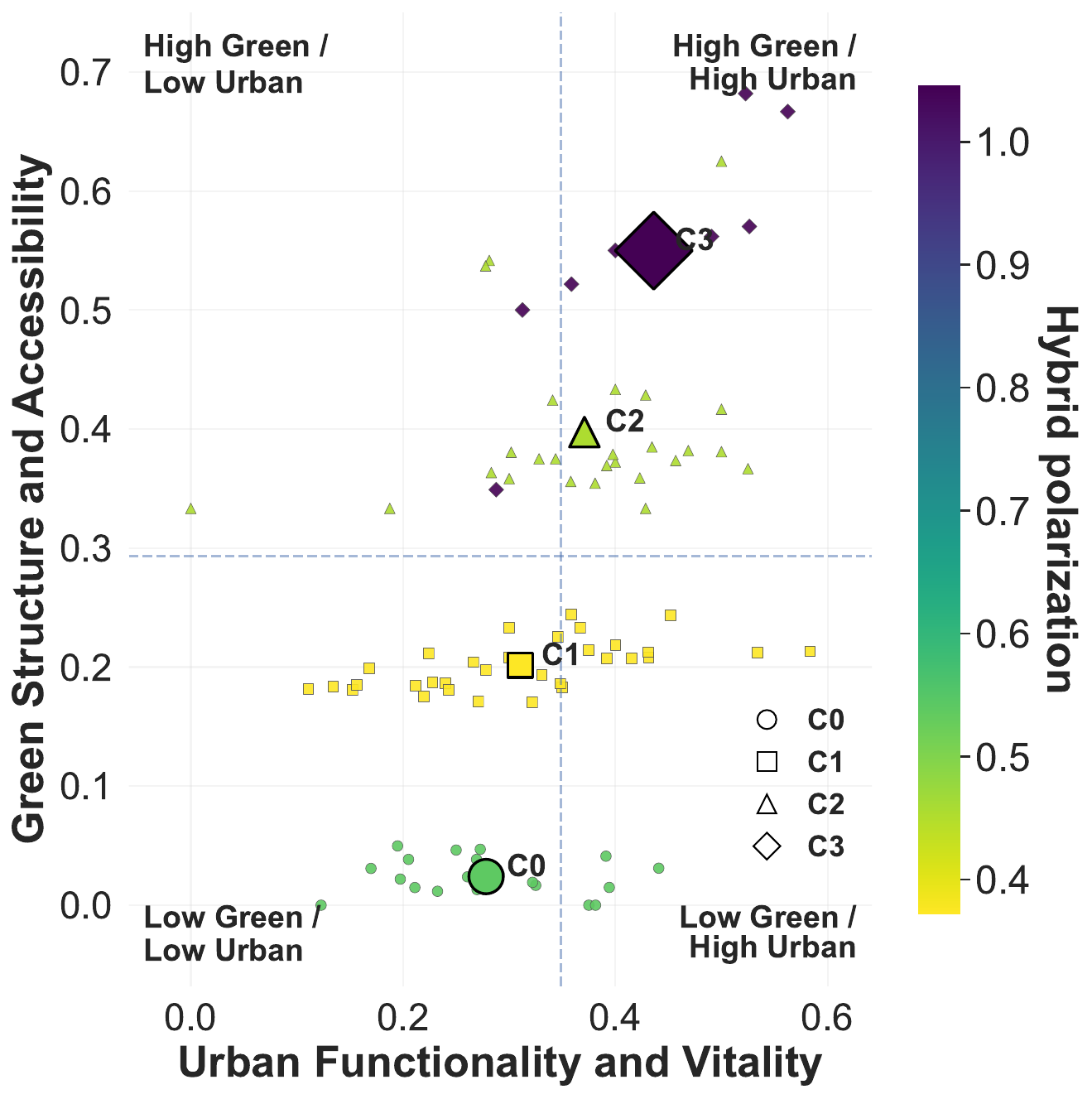}
    \caption{Polarization ($P = 1-H_{\mathrm{norm}}$).}
    \label{fig:gardner_po}
\end{subfigure}
\begin{subfigure}[b]{0.32\linewidth}
    \centering
    \includegraphics[width=\linewidth]{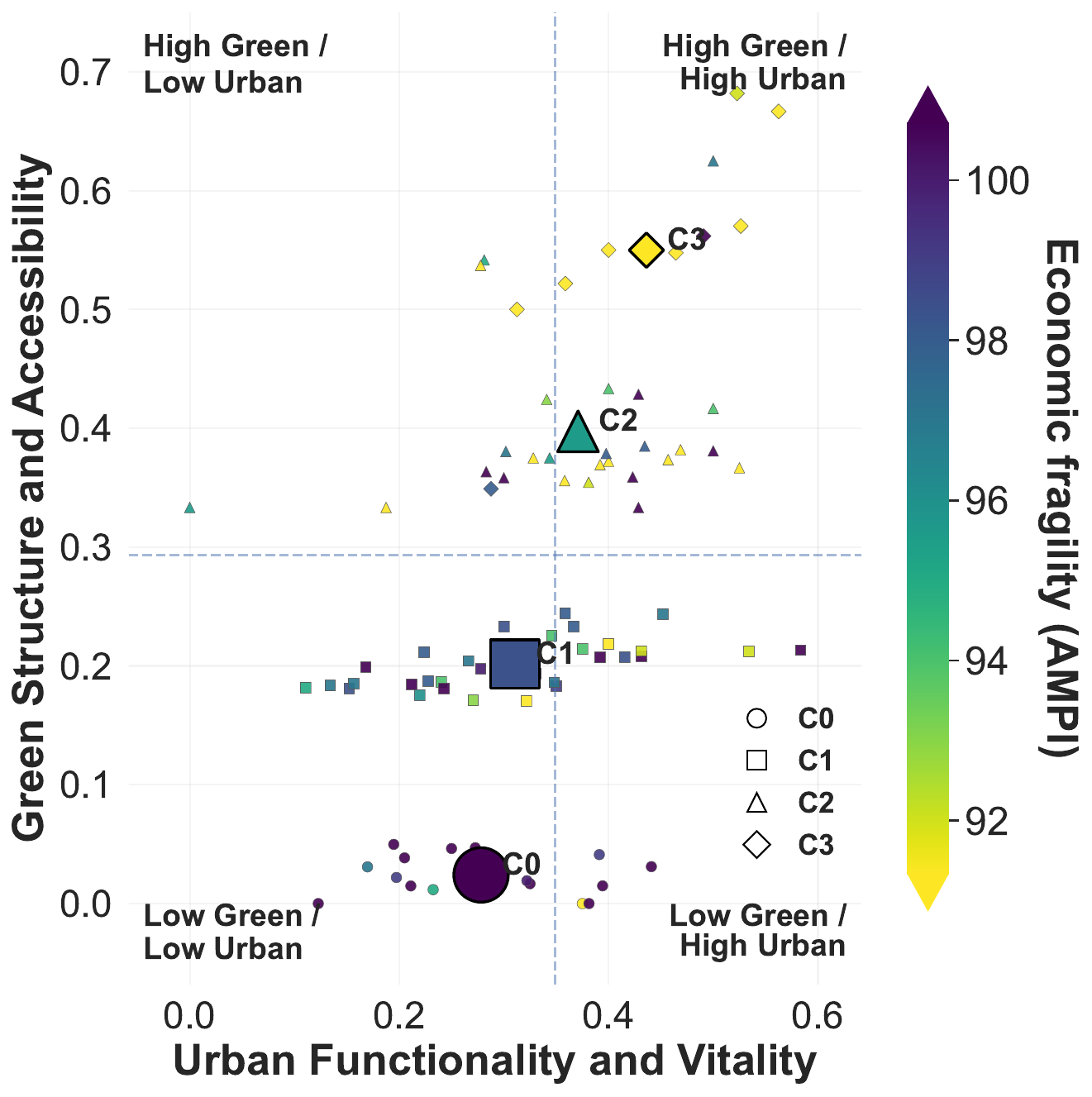}
    \caption{Economic fragility.}
    \label{fig:gardner_econ}
\end{subfigure}
\begin{subfigure}[b]{0.32\linewidth}
    \centering
    \includegraphics[width=\linewidth]{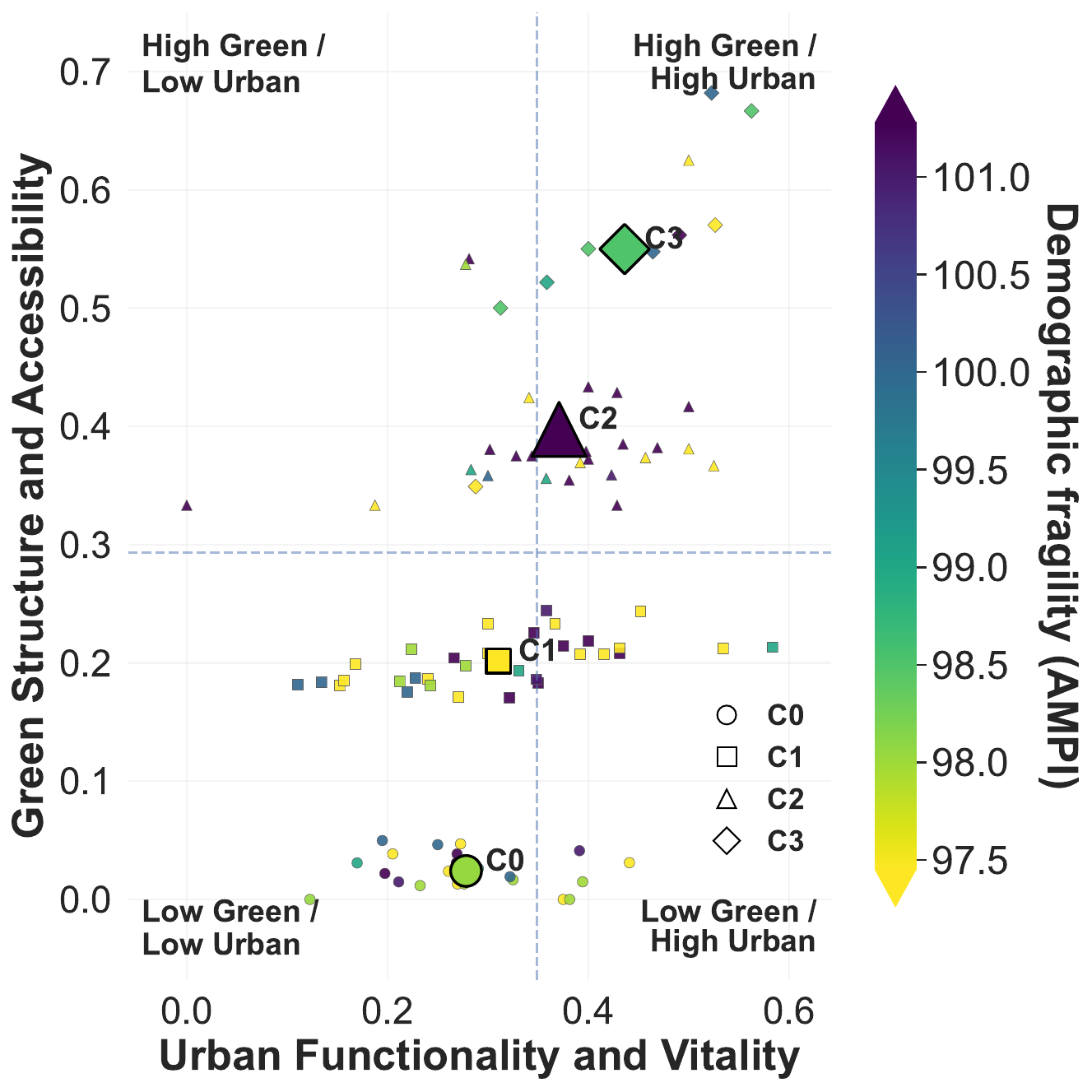}
    \caption{Demographic fragility.}
    \label{fig:gardner_demo}
\end{subfigure}
\begin{subfigure}[b]{0.32\linewidth}
    \centering
    \includegraphics[width=\linewidth]{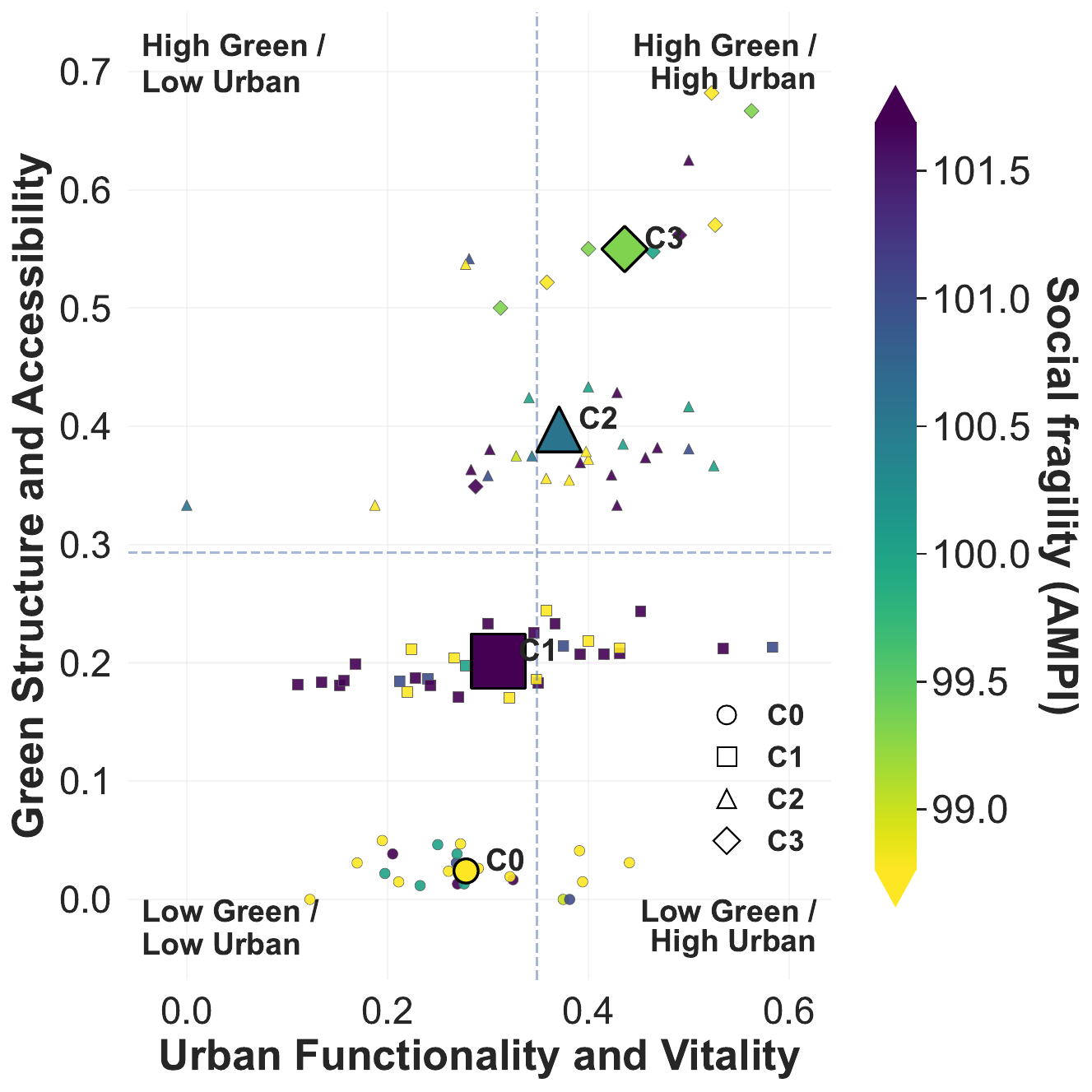}
    \caption{Social fragility.}
    \label{fig:gardner_soc}
\end{subfigure}
\begin{subfigure}[b]{0.32\linewidth}
    \centering
    \includegraphics[width=\linewidth]{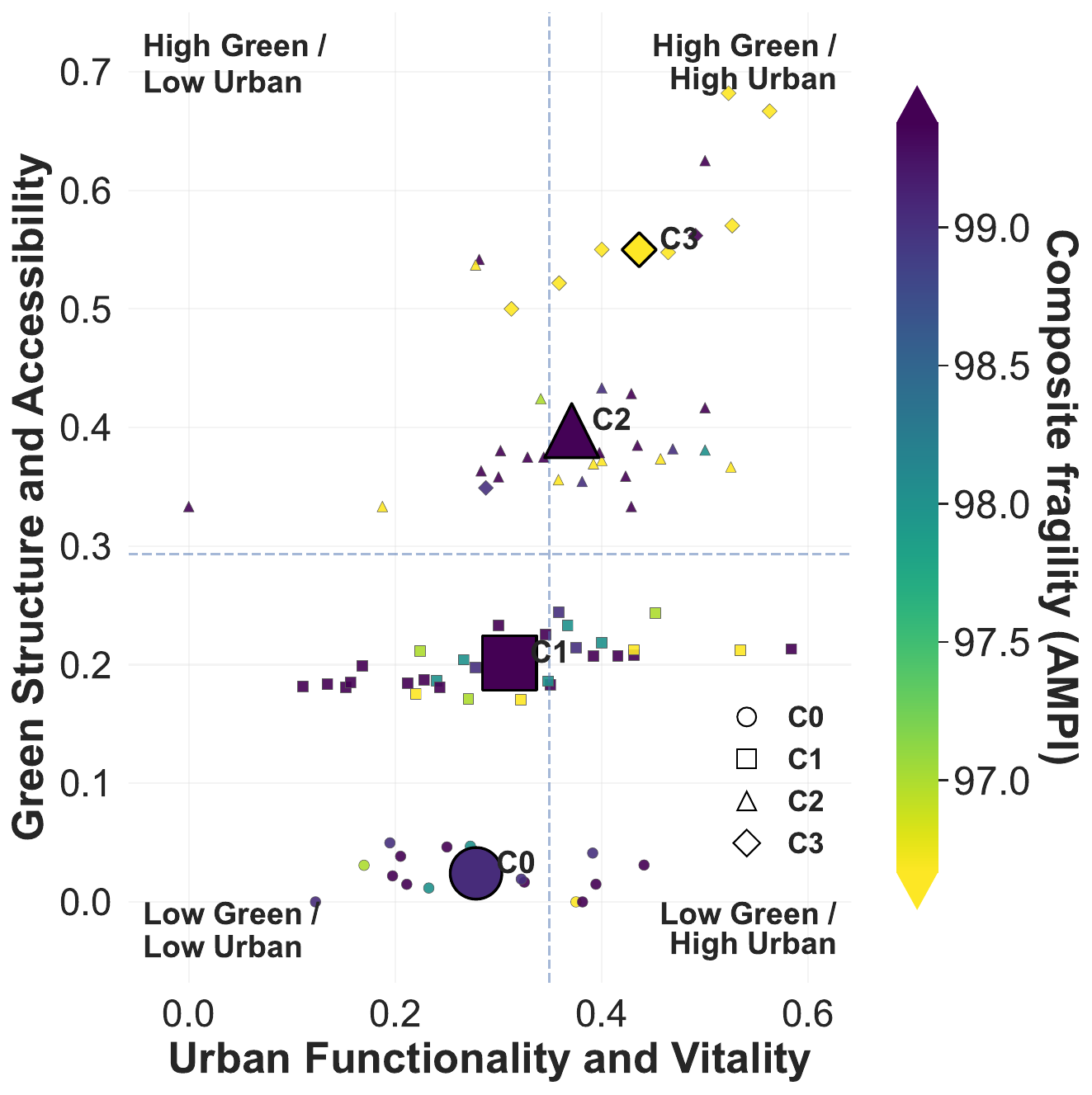}
    \caption{Composite fragility.}
    \label{fig:gardner_compl}
\end{subfigure}
\caption{Gartner-like quadrant of statistical areas (Bologna). The urban index (x-axis) and green index (y-axis) summarize service-related and green-related attributes, respectively. Large markers denote cluster centroids (\textbf{C0–C3})\rev{, and marker shape encodes cluster membership}. Marker size and color encode, respectively, polarization (panel a) and different dimensions of fragility (panels b–e). \rev{In panels (b)–(e) each small marker is a statistical area coloured by its own value of the index, while centroid colour and size encode the cluster mean; colour scales report the original AMPI values.} Dashed lines mark cross-cluster means.}
\label{fig:gardner_full}
\end{figure}

Implementing both diagnostics, the elbow criterion on within-cluster sum of squares and the average silhouette coefficient, indicated that \textit{four clusters} offer a suitable balance between model complexity and cluster quality. The resulting clusters are illustrated in Figure S7 in the SM.
We examined the distributions of the input variables across clusters (Figs.~\ref{fig:violins_group1}–\ref{fig:violins_group3}). These plots reveal ordered patterns across clusters, where several categories become increasingly prevalent in higher-numbered clusters, particularly those related to green availability, outdoor leisure, arts and cultural venues, and transportation (Figs.~\ref{fig:violins_group1} and \ref{fig:violins_group2}).
Conversely, other categories display an opposite trend, being more prevalent in the lower-numbered clusters (Fig.~\ref{fig:violins_group3}).
\rev{To verify that this distinction is a property of the partition rather than an interpretive overlay, we re-estimated the clustering on the two subsets of input variables separately and compared each resulting partition with the full one through the Adjusted Rand Index. Re-clustering on the AGORA variables alone reproduces the full partition almost exactly ($ARI = 0.94$), whereas re-clustering on the vocational categories alone does not ($ARI = 0.11$). Both values are stable across twenty random initialisations. The partition is therefore determined by the AGORA variables, while the vocational ones differentiate areas without structuring the grouping.}
To further synthesize these patterns, we constructed a Gartner-like quadrant (Fig.~\ref{fig:gardner_po}), an interpretive two-axis framework inspired by the comparative evaluation model originally proposed by Gartner Inc.~\citep{Gartner2023}. \rev{The two axes split the AGORA variables identified above into their green and urban components.} The {green index} aggregates the three binary green-infrastructure indicators together with Level-1 POI categories \texttt{Sports and Recreation}, \texttt{Outdoors and Recreation}, and \texttt{Landmarks and Outdoors}; the {urban index} aggregates \texttt{Community and Government}, \texttt{Travel and Transportation}, and \texttt{Arts and Entertainment}. For each territorial unit, the two axes are computed as the mean across the respective feature sets\rev{, so that the quadrant represents areas along the two dimensions that jointly determine the partition, while the vocational categories are left out}.
As a third attribute, we encoded polarization as
\begin{equation}\label{eq:pol}
    P \;=\; 1 - H_{\mathrm{norm}},
\end{equation}
where
\[
H_{\mathrm{norm}}(x) \;=\; \frac{-\sum_{i=1}^{m} p_i \log p_i}{\log m}, \quad
p_i \;=\; \frac{x_i}{\sum_{j} x_j}.
\]
This formulation builds on Shannon's definition of entropy as a measure of information diversity \citep{Shannon1948}. 
Here, $H_{\mathrm{norm}}$ captures the degree of internal diversification of each territorial unit across all features, while $P$ expresses its complement, i.e., the extent of concentration or specialization. 
Accordingly, $P$ increases when a unit's functional profile is dominated by a few feature types and decreases when the profile is balanced across many.\\
In the quadrant (Fig.~\ref{fig:gardner_po}), points form {four well-separated clouds}: clusters \textbf{C0} and \textbf{C1} lie predominantly in the \emph{low-green/low-urban} quadrant, while \textbf{C2} and \textbf{C3} occupy the \emph{high-green/high-urban} quadrant. Large markers denote cluster centroids, with marker size and color encoding polarization\rev{; marker shape encodes cluster membership}.
Notably, \textbf{C3}, the cluster with the highest green and urban scores, also {exhibits the strongest polarization}, indicating a profile concentrated in a subset of features rather than broadly diversified. \rev{Since polarization is computed on the average profile of each cluster, it is defined at the cluster level only: within each cloud the colour is therefore constant by construction and does not indicate internal homogeneity.} Spatially, \textbf{C3} is primarily located in Bologna's northern hillside areas, capturing major green lung, as well as several additional, mutually distant areas that are not strictly central. 
Cluster \textbf{C0} predominantly covers the {central} districts of the city, whereas \textbf{C1} and \textbf{C2} map to {peripheral} areas that are fairly interleaved spatially. To probe \rev{fragility} patterns more explicitly, we replaced the third variable in the Gartner-like plots with four fragility indices curated by the municipality. Specifically, we compared the unsupervised clusters against: demographic fragility, social fragility, economic fragility, and composite fragility.\\
\rev{The resulting overlays are reported in Figs.~\ref{fig:gardner_econ}, \ref{fig:gardner_demo}, \ref{fig:gardner_soc} and \ref{fig:gardner_compl}. Colour scales report the actual index values on the AMPI scale (higher values denote higher fragility). Centroid colour and size encode the cluster mean, each statistical area is coloured by its own value, and marker shape encodes cluster membership, so that each panel shows both the ordering across clusters and the dispersion within them.}\\
{Economic fragility} increases as one moves towards the \emph{low-green/low-urban} quadrant, corroborating that not only green assets but also urban services concentrate in economically stronger areas, Fig.\ref{fig:gardner_econ}. \rev{This is the only dimension along which the four clusters differ significantly ($p=0.003$), and the only one whose cluster means decline monotonically, from $100.71$ in \textbf{C0} to $91.33$ in \textbf{C3}. The gap of $9.38$ points corresponds to about $1.3$ standard deviations of the distribution across statistical areas, and the partition accounts for roughly $15\%$ of the variance of the index ($\eta^2=0.149$, the share of total variance attributable to differences between clusters rather than among areas within the same cluster).}
{Demographic fragility} (Fig.\ref{fig:gardner_demo}) peaks in \textbf{C2}, a cluster that still lies in the \emph{high-green/high-urban} quadrant (though not at the very top). Cross-referencing with the economic fragility map suggests that the most demographically fragile groups (e.g., older residents and low replacement) are {not} necessarily the poorest nor the most service- or green-deprived; conversely, the least demographically fragile areas tend to be younger, even when located in \emph{low-green/low-urban} contexts. \rev{The spread across cluster means is however small relative to the variation among areas within each cluster, and the differences do not reach significance ($p=0.116$).}
Finally, {social fragility} is highest in \textbf{C1}, which sits in the {low-green/low-urban} quadrant and is largely non-central, pointing to a \rev{fragility} configuration distinct from the central \textbf{C0} and the peripheral yet better-endowed \textbf{C2--C3}. \rev{Here the clusters differ under the rank-based test but not under the parametric one ($p=0.028$ and $p=0.086$): the dimension separates the clusters without following the green--urban gradient.}
Turning to the {composite} dimension (\texttt{frag\_compl}; Fig.~\ref{fig:gardner_compl}), we find that only one cluster, \textbf{C3} (the cluster with the highest green and urban endowments), consistently exhibits {low} composite fragility. \rev{Within \textbf{C3}, a single area, Pilastro, departs markedly from the others, with a composite value of $105$ against a cluster mean of $96.7$ and a minimum of $91$ in San Luca.} By contrast, the remaining clusters display {higher} composite scores but for different reasons: \textbf{C1} reflects a mix dominated by social--economic \rev{fragility} in \textit{low-green/low-urban} contexts; \textbf{C2} is lifted by demographic \rev{fragility} despite relatively favorable endowments; and \textbf{C0} is chiefly characterized by \emph{economic} fragility, despite its central location. \rev{The $2.38$-point advantage of \textbf{C3} over \textbf{C0} does not, however, make the four clusters statistically distinguishable on this dimension ($p=0.200$), since aggregating three axes that respond to the green--urban structure in different ways attenuates the ordering that the economic component displays on its own.}
\rev{Table~\ref{tab:variance_decomposition} makes explicit how much of each index the partition accounts for, by splitting its variance into a between-cluster share, which is what the clusters capture, and a within-cluster share, which is variation among areas assigned to the same cluster. The between share reaches $14.9\%$ for economic fragility and stays at or below $7.3\%$ elsewhere, and differences in cluster means are statistically significant only for the economic dimension ($p = 0.003$). A modest between share is expected, since the clusters are built on service and green indicators alone, without any information on fragility. Internal dispersion shows no significant differences across clusters in any dimension (Brown--Forsythe $p \geq 0.081$), so no cluster stands out as markedly more heterogeneous than the others.}
Overall, the pattern suggests that abundant green and service endowments are \rev{associated with lower fragility, most clearly along its economic dimension, while the remaining dimensions distribute across clusters according to configurations that do not follow the green--urban gradient.}
\begin{table}[htbp]
\centering
\caption{\rev{Variance of each municipal fragility index in Bologna, split into within- and between-cluster shares, with tests of difference across the four clusters. The between share equals $100\,\eta^2$. Brown--Forsythe tests the equality of within-cluster variances.}}
\label{tab:variance_decomposition}
\small
\begin{tabular}{lrrrr}
\toprule
Dimension & Within (\%) & Between (\%) & $F$ ($p$) & Brown--Forsythe $p$ \\
\midrule
Demographic & 93.4 & 6.6  & 2.030 (0.116) & 0.081 \\
Social      & 92.7 & 7.3  & 2.270 (0.086) & 0.639 \\
Economic    & 85.1 & 14.9 & 5.003 (0.003) & 0.686 \\
Composite   & 94.8 & 5.2  & 1.582 (0.200) & 0.087 \\
\bottomrule
\end{tabular}
\end{table}

\subsection{Clustering of Elementary Areas in the City of Florence}
\rev{Applying the same methodological framework to Florence, the elbow and silhouette diagnostics once again identified four clusters (\textbf{C0–C3}) as the optimal configuration, with a spatial arrangement and a functional differentiation closely mirroring those of Bologna (Figures S8 and S9 in the SM). Cluster \textbf{C0} corresponds to the central elementary areas, covering the historical and most urbanized core; \textbf{C1} and \textbf{C2} represent mixed peripheral areas, geographically interleaved across different districts and not clearly separated from one another; \textbf{C3} aggregates the greenest elementary areas, mostly situated in the hillside belt surrounding Florence, including the southern and northeastern slopes.\\
The Gartner-like quadrant, constructed with the same green and urban indices and the same polarization measure adopted for Bologna (Figure S10(a) in the SM), reproduces Bologna's configuration: all cluster centroids fall in the \textit{high-green/high-urban} and \textit{low-green/low-urban} quadrants, symmetrically arranged along the green--urban gradient. \textbf{C0} sits at the compact, highly urbanized end and \textbf{C3} at the opposite one, while \textbf{C1} and \textbf{C2} occupy intermediate positions with partially overlapping point clouds, indicating only moderate functional differentiation. The reproduction of this structure in a second city confirms the transferability and structural consistency of the framework.\\
We now extend the quadrant-based interpretation by introducing a third dimension capturing socio-demographic fragility. Differently from Bologna, the four clusters in Florence reveal a partially distinct {fragility} configuration. Economic fragility remains highest in \textbf{C0}, confirming its structural disadvantage in both cities (Figure S10(b) in the SM). Demographic fragility is concentrated in \textbf{C0} and \textbf{C1}, whereas \textbf{C2} {shows no comparable demographic burden} (Figure S10(c) in the SM). Social fragility, instead, is more uniformly distributed and affects \textbf{C0}, \textbf{C1}, and \textbf{C2}, with \textbf{C0} emerging as the most socially {fragile} cluster, whereas in Bologna this role was played by \textbf{C1} (Figure S10(d) in the SM). Consistently, composite fragility increases as the cluster index decreases, indicating that the {least fragile} areas are those with {higher} green endowment and {greater} service accessibility (Figure S10(e) in the SM). \textbf{C3} records the lowest values on every fragility dimension.}

\subsection{AGORA Alignment as an Urban Ordering Mechanism}
\label{sec:agora_alignment}
\begin{figure}[t]
    \centering
    \includegraphics[width=0.75\textwidth]{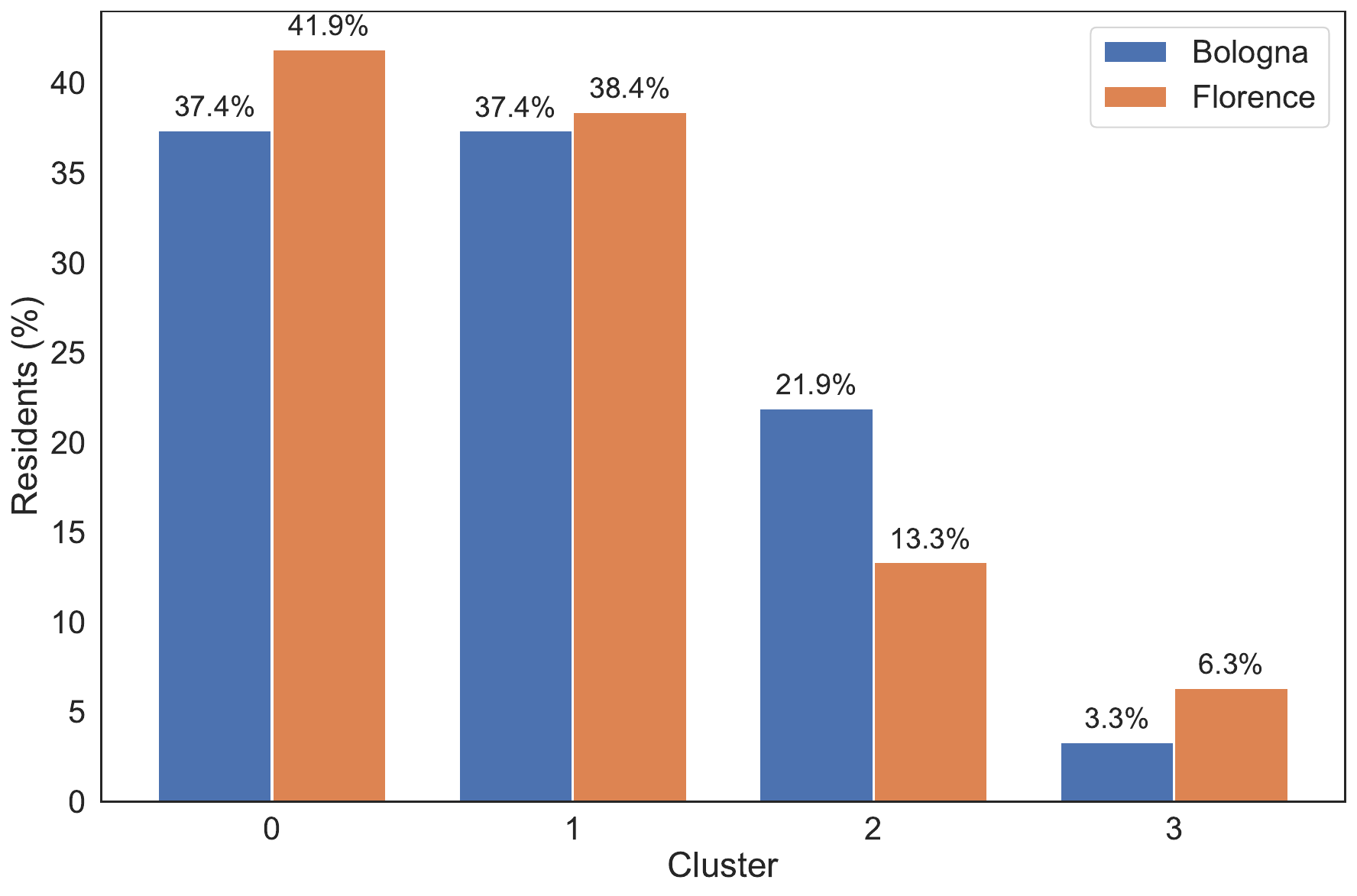}
    \caption{Share of residents by cluster in Bologna and Florence. 
    The non-fragile cluster (Cluster~3) accounts for only 3.3\% of residents in Bologna and 6.3\% in Florence, 
    highlighting how fully non-fragile contexts remain rare in both cities.}
    \label{fig:residenti-cluster}
\end{figure}

\rev{In both cities, clusters with higher endowments of parks and tree canopy, outdoor and recreational amenities, community and governmental services, and public transportation tend to display lower fragility, while those in the low-green/low-urban quadrant concentrate the highest levels (Figs.~\ref{fig:BO_clusters_full} and~\ref{fig:gardner_compl}; Figs.~S9 and~S10(e) in the SM). The association is clearest for economic fragility, whereas demographic fragility follows a partly distinct pattern in Bologna. We refer to the joint presence of these infrastructures as \emph{AGORA}: Arts and cultural life, Green and recreational environments, and Open Routes for Access and mobility. To measure how consistently fragility decreases along this gradient, we introduce an \emph{AGORA Alignment} score (AA).} For each cluster \(k\), we define a shelter score
\[
    s_k \;=\; \frac{u_k + g_k}{2},
\]
where \(u_k\) and \(g_k\) denote the centroid coordinates on the urban and green axes, respectively. Let \(f_k\) be the average fragility of cluster \(k\). The AGORA Alignment score is then computed as
\[
    \mathrm{AA} \;=\; -\,\rho(s_k,\, f_k),
\]
where \(\rho(\cdot,\cdot)\) is Spearman's rank correlation coefficient.
\rev{An AA of 1 indicates that fragility decreases monotonically as AGORA endowment grows. The score is computed on cluster centroids, which represent archetypal urban configurations, so that the largest clusters are not over-weighted; it is therefore a descriptive indicator rather than an inferential statistic. Consistently with Table~\ref{tab:variance_decomposition}, where fragility varies mostly within clusters and cluster means differ significantly only for the economic dimension, AA captures an ordering of typical configurations rather than a partition of fragility itself.\\
Florence shows a complete ordering (AA $= 1.00$): composite fragility decreases at every step of the AGORA gradient. Bologna shows a strong but incomplete ordering (AA $= 0.80$), due to a single rank inversion between \textbf{C1} and \textbf{C2}: although \textbf{C2} ranks higher on the shelter score $s_k$, its composite fragility exceeds that of \textbf{C1}, driven by its demographic component (Fig.~\ref{fig:gardner_demo}), while \textbf{C0} and \textbf{C3} retain the expected extreme positions. Under economic fragility alone, Bologna also attains AA $= 1.00$.} Based on these findings, we define:
\begin{definition}
\textbf{AGORA-Aligned cities} are urban systems where fragility decreases as access to jointly distributed AGORA infrastructures increases. These infrastructures encompass arts and cultural-community services, green and recreational spaces, and open mobility routes.
\end{definition}
\rev{The reach of this configuration is, however, limited: only 3.3\% of residents in Bologna and 6.3\% in Florence live in the least fragile cluster, the only one with full AGORA endowment, while the share of residents grows in the more fragile clusters (Fig.~\ref{fig:residenti-cluster}). This gap between alignment and coverage points to the need to extend AGORA infrastructures across more of the urban fabric, without altering the functional vocations of individual neighbourhoods.}
\section{Discussion}
\label{sec:discussion}
\rev{In this study, drawing on a multidimensional dataset that integrates green infrastructure with both essential and non-essential urban services in Bologna and Florence, we show that urban fragility is lower in urban configurations where green spaces, community services, cultural and recreational amenities, and open mobility infrastructures are jointly accessible.}
We define an \emph{AGORA Alignment score (AA)} that measures how consistently a city exhibits this ordering, whereby \rev{fragility} decreases as access to AGORA infrastructures increases, and refer to cities that follow this pattern as \emph{AGORA-Aligned}.
\rev{The score is computed on a composite measure integrating economic, social and demographic fragility and operates at the level of cluster archetypes, capturing the ordering of typical urban configurations along the green--urban gradient. This level of analysis is not incidental: most of the variance of each fragility index lies within rather than between clusters (Table~\ref{tab:variance_decomposition}), so the ordering AA measures is a property of urban configurations rather than of individual areas, and areas sharing a configuration may still differ appreciably in their fragility.}
\rev{Florence is fully AGORA-Aligned (AA = 1.00), with composite fragility decreasing at every step of the gradient, while Bologna shows strong but incomplete alignment (AA = 0.80).}
While this holistic formulation reflects the multidimensional nature of \rev{fragility}, \rev{it also attenuates the ordering, since it aggregates dimensions that respond to the green--urban structure in different ways:} Bologna \rev{attains} full AGORA alignment (AA = 1.00) \rev{when} only economic fragility \rev{is} considered\rev{, the dimension along which the clusters differ significantly (Table~\ref{tab:variance_decomposition}), whereas demographic and social fragility peak in intermediate clusters}.
\rev{The association between AGORA availability and lower fragility is therefore not uniform across dimensions.}
For example, in Bologna, demographic fragility peaks in Cluster~2 (Figure~\ref{fig:gardner_demo}), which is located within
the high green/high urban quadrant, suggesting that the most \rev{fragile} groups from a demographic point of view are not necessarily the poorest or those most deprived of services.\\

\rev{Service categories such as Retail, Dining and Drinking, Health and Medicine, and Business and Professional Services instead show wide and overlapping distributions across clusters and are most abundant in the central cluster, reflecting centrality and market demand rather than ordering clusters by fragility.}
Taken together, the analysis of Bologna and Florence reveals a dual organization of urban space: a vocational layer, which describes how areas operate, and an AGORA layer, which signals where \rev{fragility} arises. The coexistence of these layers indicates that similar vocational profiles can correspond to very different levels of infrastructural accessibility and, consequently, to different fragility conditions.
Vocational services primarily support the routine functioning of urban life and respond dynamically to local demand and market conditions. As a result, they tend to reorganize spatially into specialized and differentiated clusters, giving rise to distinct vocational profiles across neighborhoods. Their relative flexibility and substitutability allow similar functional roles to be fulfilled through different local configurations, weakening their systematic association with \rev{fragility} patterns.
In contrast, AGORA infrastructures
operate as collective and cumulative resources whose benefits depend on coordinated spatial accessibility and long-term public investment:
(i)~green spaces \rev{support psychological restoration} \citep{kaplan2001nature} \rev{ and provide broader health and climate-regulation benefits \citep{hartig2014nature}};
(ii)~cultural and community services support social belonging and personal growth \citep{klinenberg2018palaces};
(iii)~open mobility expands the accessibility to space \citep{bertolini2012integrating}.
\rev{Together, these components may contribute to the conditions underlying fragility patterns, rather than merely reflecting local functional specialisations.}
The composite fragility index employed in our analysis aggregates demographic, social, and economic \rev{disadvantages} (Table~S1 in the SM), capturing what Evans and Kantrowitz (2002) \citep{evans2002socioeconomic} conceptualize as cumulative deprivation: the non-additive amplification of disadvantage when multiple stressors co-occur.
\rev{Within this framework, AGORA deprivation may act as an additional stressor: areas in Clusters~0--2 lack integrated access to green space, cultural amenities and mobility infrastructure and display elevated fragility, albeit along different dimensions in each cluster. Whether the absence of AGORA infrastructures compounds other forms of disadvantage, rather than merely co-occurring with them, remains to be tested.}\\
\rev{These mechanisms are drawn from the literature and are not tested here: our design identifies where fragility and AGORA deprivation co-occur, but cannot establish why. Two alternatives must be distinguished. AGORA infrastructures may reduce disadvantage through the mechanisms above, or fragile households may concentrate where housing is cheaper and AGORA endowment is scarce, through residential sorting. Since restoration and belonging are individual processes, and associations observed across areas cannot establish mechanisms operating on residents, the first pathway would have to be tested at the individual level, by surveying residents across the four clusters on perceived restorativeness, sense of neighbourhood belonging and perceived accessibility of daily destinations, and testing whether these mediate the association between cluster membership and the outcomes that make up fragility, such as health and social isolation. The cluster classification provides the sampling frame for such a study. Separating this pathway from residential sorting requires longitudinal designs: greening or pedestrianisation schemes could serve as natural experiments, comparing changes in outcomes among long-term residents of treated and comparable untreated areas.}\\

\rev{The polarization of the least fragile cluster qualifies this further. Its profile is concentrated in the AGORA components, while the services of everyday life (retail, dining, health and professional services) are comparatively scarce, as the distributions of the input variables across clusters show. The configuration associated with the lowest fragility is therefore a specialised, largely residential one, confined to a morphologically privileged belt and reaching only $3.3\%$ of residents in Bologna and $6.3\%$ in Florence. Polarization is used here in a functional sense, as the concentration of a unit's service profile, and is distinct from social polarization; the two are nonetheless connected, since spatial concentration is selective and context-dependent and feeds back into the mechanisms producing inequality \citep{Pratschke2012}. The least fragile cluster is therefore not a template to be replicated: its endowment rests on morphological conditions that most of the urban fabric does not possess.}
\rev{That this configuration should be so rare is not incidental.} These infrastructures are not systematically guaranteed by routine urban service provision and market-driven dynamics. As a result, they tend to concentrate in areas characterized by greater purchasing power, stronger political advocacy, and higher levels of social capital.
\rev{This may produce a form of deprivation in AGORA access that is difficult to offset: when AGORA infrastructures are absent from a neighbourhood, few alternatives can fill the gap, whereas vocational services can be more easily substituted, replaced, and reconfigured.}
These substitutes maintain a baseline level of service provision that mitigates the link between vocational infrastructure and fragility. 

\rev{The relevant objective is therefore to raise the missing AGORA components in areas where vocational services are already well provided and where the large majority of residents live: expanding public access to green areas, community spaces and cultural life should become a transversal planning objective, not a prerogative of already advantaged neighbourhoods.}
\rev{Such expansion, however, carries a risk highlighted by the environmental justice literature on green gentrification.}
Studies show that investments in urban greening, parks,
and tree planting often trigger
property value increases that subsequently displace the very low-income and marginalized communities the investments were intended to benefit \citep{checker2011wiped}, \citep{anguelovski2019new}.
\rev{Comparative evidence shows that this outcome is not inevitable but varies from city to city, depending on housing affordability and anti-displacement policies \citep{Anguelovski2022green}: AGORA expansion should thus be paired with such instruments, not pursued in isolation. In our results the concern is prospective, since the best-endowed cluster corresponds in both cities to hillside belts whose greenery is morphological and long-standing, not the product of recent interventions.} A one-size-fits-all approach to AGORA provision is therefore unlikely to succeed.
Different population groups require distinct configurations of civic infrastructure to achieve wellbeing \citep{giles2016city}.
Urban planners and local policy makers should engage in participatory processes to identify which combinations of green space, cultural amenities, and mobility options best serve local communities.
\rev{Participation, however, tends to reproduce the inequalities it is meant to address, since residents with more time, resources and institutional familiarity are over-represented, and consultation without decision-making power offers little influence over outcomes \citep{arnstein1969ladder}. The fragility indices used here can help counter this bias: they identify the areas and population segments to be prioritised, allow recruitment targets to be set in proportion to their share of residents, and provide a benchmark for checking whether those who take part reflect the population concerned. Reaching these groups also requires working through intermediaries they already trust, such as social services, schools and local associations, and adapting formats, times and languages to their constraints.}
\rev{This approach can preserve the functional vocations of neighbourhoods while extending AGORA access across the urban fabric.}\\

The scope of these findings remains bounded by the empirical context of two Italian cities and by the use of open data sources. This choice enables transparency and broad methodological replicability, since satellite imagery and volunteered geographic information on urban services are publicly accessible and do not depend on proprietary datasets. However, open spatial data may vary in completeness and positional accuracy across urban environments, particularly when information relies on user contributions. A similar consideration applies to the fragility indicators: while Bologna and Florence offer open demographic, social and economic data at a suitable resolution, comparable indicators are not universally available across cities, potentially constraining the extension of this approach to other contexts.
\rev{This constraint is likely to be most binding in cities of the Global South, where the informality of much settlement, the uneven coverage of volunteered and administrative sources, and the different meanings attached to open and green space \citep{rigolon2018access} imply that the procedure is transferable but the indicator set must be re-specified locally rather than imported.}
\rev{A further question concerns the role of urban scale. Bologna and Florence are comparable in population size, and this comparability is what makes the reproduction of the same cluster structure informative, since it emerges under equivalent conditions. It does not, however, establish that the pattern is scale-invariant. Since the density and diversity of urban amenities scale systematically with population \citep{alessandretti2020scales,Bettencourt2013}, larger cities may sustain multiple, spatially distinct AGORA configurations rather than the single centre--periphery gradient we observe, while smaller ones may lack the critical mass required for the joint provision of cultural, green and mobility infrastructures to emerge at all. Testing whether AA varies systematically with city size, and whether the same three infrastructural components recur across scales, is a direct extension of the present work.}
A key question for future work is whether the same categories identified here, green spaces, cultural-community services, and open mobility, consistently mark areas of lower fragility in other contexts, or whether different urban systems exhibit alternative sets of infrastructures \rev{associated with lower fragility. Addressing this question would test the generality of AGORA and could reveal whether multiple civic models are associated with a lower spatial concentration of fragility.}

\section*{Data Availability Statement}
All data sources are described and referenced in the manuscript and are publicly available at the cited sources.
 \section*{Acknowledgments}
 This work was carried out within the framework of the TALEA Green Cells - Leading the Green Transition project (EUI02-064). The authors also wish to thank Gianni Dugheri and Emanuele Gheri from the Municipality of Florence for their availability and support in providing access to the data used for the Florence case study, and Serena Orlandi from the Department of Architecture at the University of Bologna for the insightful discussions that contributed to the refinement of this work.
\bibliography{biblio}

\end{document}